\documentclass[11pt,preprint]{aastex}

\shortauthors{Boden et al.}
\shorttitle{V773~Tau~B}

\begin{document}

\title{A Surprising Dynamical Mass for V773~Tau~B}

\author{
  Andrew F.~Boden\altaffilmark{1},
  Guillermo Torres\altaffilmark{2},
  Gaspard Duch\^ene\altaffilmark{3,4},
  Quinn Konopacky\altaffilmark{5,6}, \\
  A.M.~Ghez\altaffilmark{6},
  Rosa M.~Torres\altaffilmark{7,9},
  Laurent Loinard\altaffilmark{8,9}
}

\altaffiltext{1}{Division of Physics, Mathematics, and Astronomy,
  California Institute of Technology, MS 11-17, Pasadena, CA 91125}

\altaffiltext{2}{Harvard-Smithsonian Center for Astrophysics, 60
Garden St., Cambridge MA 02138}

\altaffiltext{3}{Division of Astronomy and Astrophysics, University of California, Berkeley, CA}

\altaffiltext{4}{UJF-Grenoble 1 / CNRS-INSU, Institut de Plan\'etologie et 
d'Astrophysique de Grenoble (IPAG) UMR 5274,
Grenoble, F-38041, France}

\altaffiltext{5}{Lawrence Livermore National Laboratory, 7000 East Avenue, Livermore, CA 94550}

\altaffiltext{6}{Department of Physics \& Astronomy, UCLA, Los Angeles, CA 90095-1562}

\altaffiltext{7}{Argelander-Institut f\"ur Astronomie,
Universit\"at Bonn, Auf dem H\"ugel 71, 53121 Bonn, Germany}

\altaffiltext{8}{Max-Planck-Institut f\"ur Radioastronomie, Auf dem H\"ugel 69, 53121 Bonn, Germany}

\altaffiltext{9}{Centro de Radiostronom\'{\i}a y Astrof\'{\i}sica,
Universidad Nacional Aut\'onoma de M\'exico, Apartado Postal 72--3
(Xangari), 58089 Morelia, Michoac\'an, M\'exico}

\keywords{binaries: spectroscopic --- stars: circumstellar matter
  --- stars: pre-main sequence --- stars: individual (V773~Tau)}

\slugcomment{v0.9 Accepted Version 29 November 2011}

\begin{abstract}

We report on new high-resolution imaging and spectroscopy on the
multiple T Tauri star system V773 Tau over the 2003 -- 2009 period.
With these data we derive relative astrometry, photometry between the
A and B components, and radial velocity (RV) of the A-subsystem
components.  Combining these new data with previously published
astrometry and RVs, we update the relative A-B orbit model.  This
updated orbit model, the known system distance, and A subsystem
parameters yields a dynamical mass for the B component for the first
time.  Remarkably the derived B dynamical mass is in the range of 1.7
-- 3.0 M$_\sun$.  This is much higher than previous estimates, and
suggests that like A, B is also a multiple stellar system.

Among these data, spatially-resolved spectroscopy provide new insight
into the nature of the B component.  Similar to A, these near-IR
spectra indicate that the dominant source in B is of mid-K spectral
type.  If B is in fact a multiple star system as suggested by the
dynamical mass estimate, the simplest assumption is that B is composed
of similar $\sim$ 1.2 M$_\sun$ PMS stars in a close ($<$ 1 AU) binary
system.  This inference is supported by line-shape changes in near-IR
spectroscopy of B, tentatively interpreted as changing RV among
components in V773~Tau~B.

Relative photometry indicate that B is highly variable in the near-IR.
The most likely explanation for this variability is circum-B material
resulting in variable line-of-sight extinction.  The distribution of
this material must be significantly affected by both the putative B
multiplicity, and the A-B orbit.

\end{abstract}

\section{Introduction}

\objectname[V773 Tau]{V773~Tau} (HDE~283447, HBC~367) is among the
most remarkable pre-main sequence (PMS) stellar systems presently
known.  V773 Tau exhibits a broad variety of observable properties
traceable to its PMS status, circumstellar material, and multiplicity.
V773~Tau was first identified as a T~Tauri star by \citet{Rydgren1976}
based on H$\alpha$ and Ca II H and K emission, high lithium abundance,
photometric variability, and K2 spectral type.  The object presented
an enigmatic mixture of classical (CTTS) and weak-lined T Tauri (WTTS)
characteristics until it became clear there are multiple components;
V773~Tau was resolved as a visual binary independently by
\citet{Ghez1993} and \citet{Leinert1993} (with visual components
designated here as A and B).  \citet{Martin1994} first suggested, and
\citet{Welty1995} established A as a short-period (51-d) double-lined
spectroscopic binary (SB2).  In filled-aperture high-angular resolution
studies, \citet[herein D2003]{Duchene2003} and \citet{Woitas2003}
independently identified an additional, ``infrared'' component in the
system (herein designated C -- note D2003 use an alternate component
notation), making V773~Tau at least a compact quadruple system with no
fewer than four stars within roughly 100 AU.

The SB2 A subsystem was resolved by near-IR \citep[herein
  B2007]{Boden2007} and radio \citep[B2007]{Phillips1996}
interferometry, allowing physical orbit reconstruction and component
dynamical mass estimates (B2007).  Further, the variable radio
emission and resolved morphology has been studied to infer
magnetospheric interaction and complicated emission topology
\citep[and references]{Massi2002,Massi2006,Massi2008}.  Finally in a
companion paper to this one, the VLBA astrometric study of
\citet[][herein T2011]{Torres2011} has refined the V773 Tau A orbit
and revisited the system distance by both orbital and trigonometric
parallax, indicating that the system is at 132.8 $\pm$ 2.4 pc
\citep[an earlier VLBI trigonometric parallax by][had estimated 148.4
  $\pm$ 5.5 pc, while the B2007 orbital analysis yielded 136.2 $\pm$
  3.7 pc]{Lestrade1999}.

Since the D2003 results and modeling, near-IR imaging and
spectroscopic monitoring of V773 Tau (combined with B2007 results on
A) have yielded important additional clues to the nature of the B
component.  Here we will report on this continued monitoring, and what
data taken over the past few years imply about the B component of this
remarkable system.

\section{Observations}
\label{sec:observations}

We report on three types of new observations of V773 Tau.  New
resolved adaptive optics (AO)-corrected imaging from various
telescopes provides relative astrometry and photometry among the V773
Tau components.  New spectroscopic observations yielding radial
velocity (RV) measurements of the two A constituents (following data
presented in B2007) extend the time baseline for both A and A-B orbit
modeling.  Finally, AO-resolved near-IR spectroscopy separates the
emission from A and B for the first time, and appears to show
photospheric lines from multiple stellar components for both A and B.
We will discuss each data set in turn.

\subsection{Imaging}
\label{sec:imaging}

\begin{figure}[t]
\epsscale{0.6}
\plotone{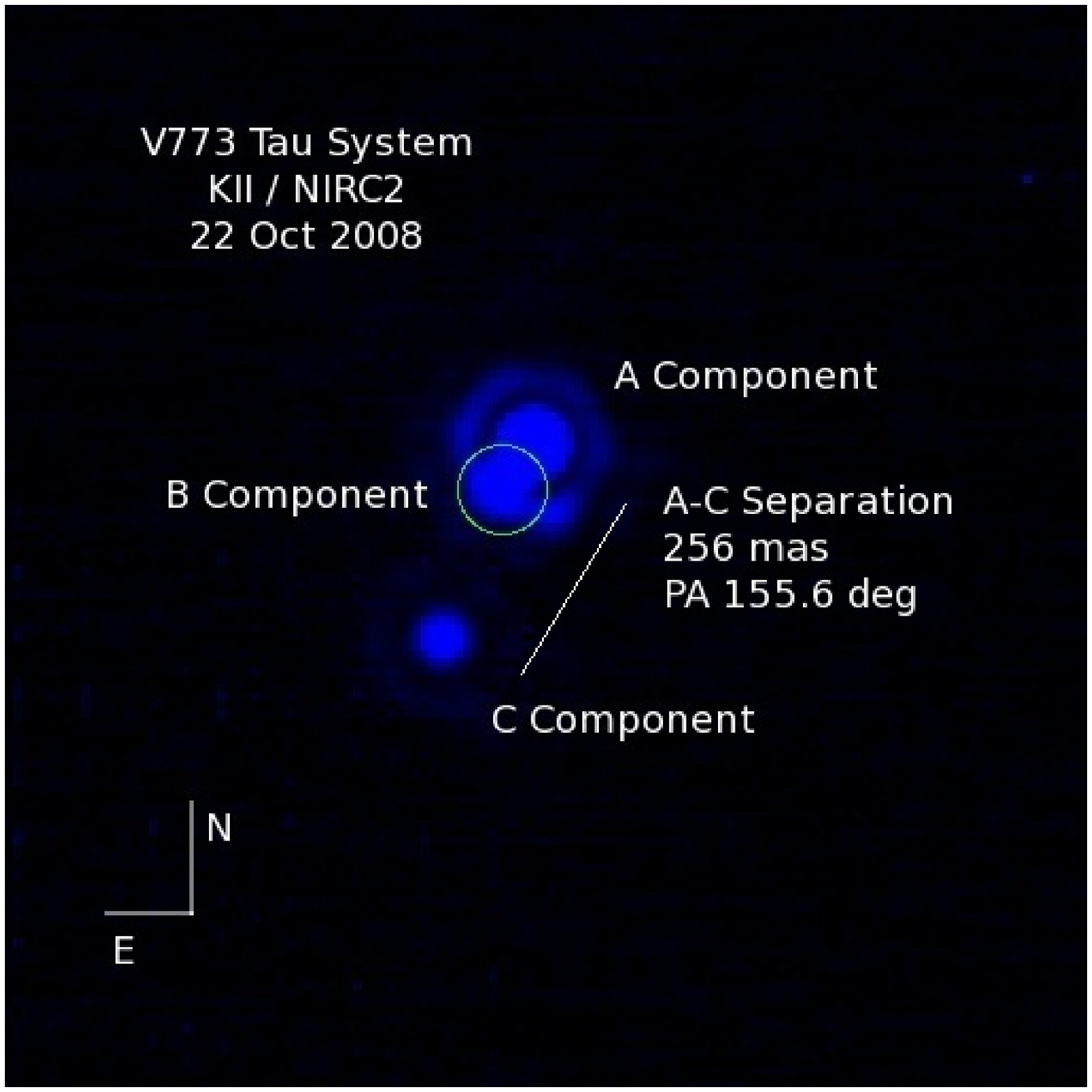}
\caption{Keck II Adaptive Optics Image of V773 Tau from 2008 Oct 22.
  The image is stretched to enhance low-level detail, so in addition
  to the A, B, and C visual components, a hexagonal diffraction ring
  around the bright A component is apparent.  At the epoch of this
  image the apparent A-B separation is roughly 50 mas at a position
  angle of 150 deg, and the apparent A-C separation is roughly 260 mas
  at a position angle of 156 deg.
\label{fig:v773TauImage}}
\end{figure}

Since the time of D2003 we have continued to monitor the V773~Tau
system in imaging.  As typical apparent spacings between the outer
components in the system are well under an arcsecond, adaptive
optics-corrected imaging is required to resolve these visual
components.  Figure~\ref{fig:v773TauImage} illustrates a
spatially-resolved K-band image of V773~Tau from 2008 Oct 22 obtained
with the facility AO system \citep{Wizinowich2000} and the NIRC2
instrument (PI: K. Matthews) on the Keck II telescope in Mauna Kea HI.

\begin{figure}[t]
\epsscale{1.1}
\plottwo{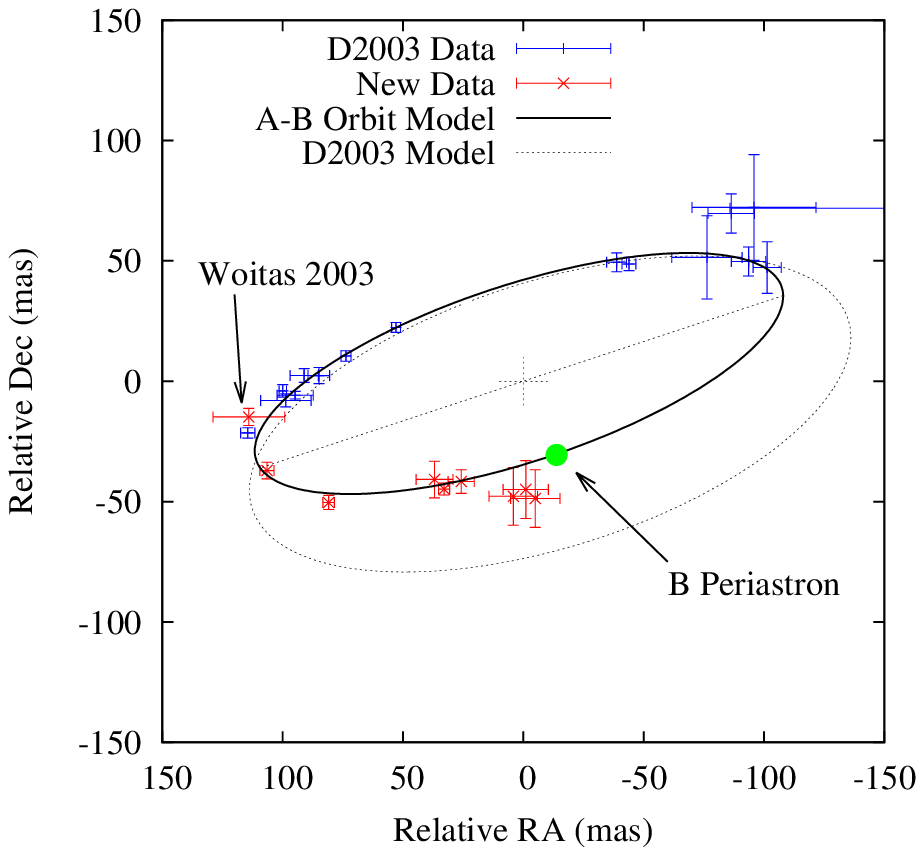}{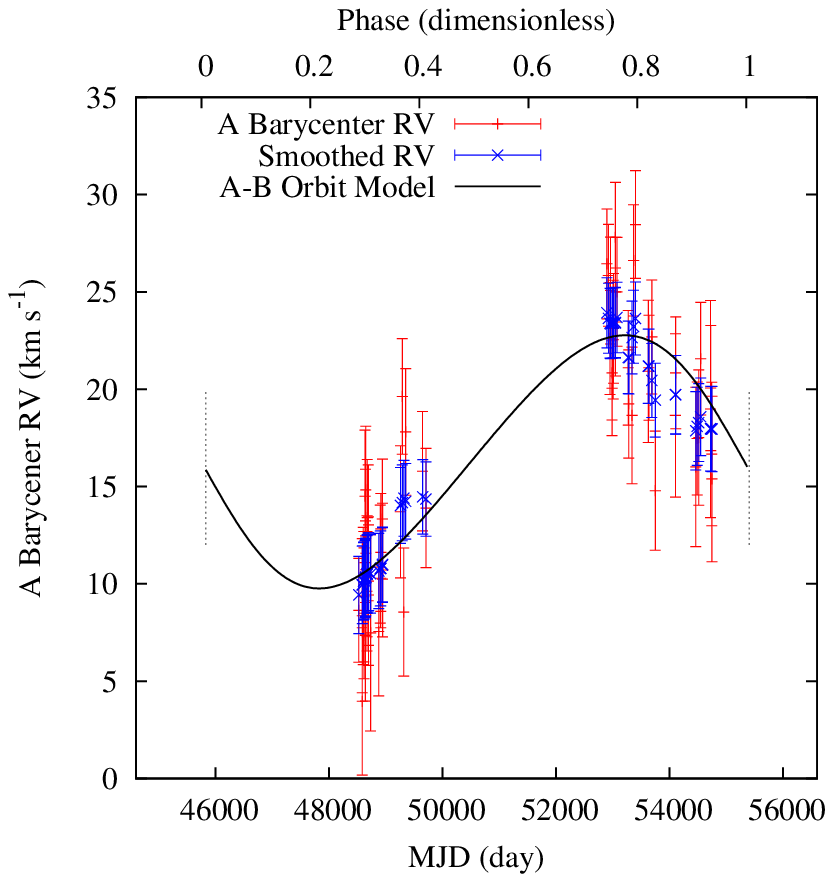}
\caption{Astrometric and Radial Velocity Data and Orbit Modeling for
  V773 Tau A-B.  Left: we depict A-B relative astrometry (D2003 data
  in blue, new data in red -- Table~\ref{tab:observations}) and the
  best-fit A-B astrometric orbit models of D2003 and present results
  (Table~\ref{tab:ABorbit}).  Right: derived A-barycenter RV
  (Table~\ref{tab:RVdata}) and our best-fit RV orbit model.  Shown in
  red are the derived A-barycenter RV as used in the orbit modeling,
  and shown in blue are a version of these same data smoothed over the
  A-subsystem orbit period (51.1d; B2007).  The smoothed data are
  rendered here to illustrate how these data pertain to the
  longer-period A-B orbit, they are not used as input to the orbit
  modeling.
\label{fig:ABorbit}}
\end{figure}

Table \ref{tab:observations} summarizes the eight new resolved imaging
data sets available to us since D2003, including derived relative
astrometry and relative photometry.  Seven of these data sets were
obtained at Keck with NIRC2, and one was produced by the VLT/NACO
system \citep{Rousset2003,Lenzen2003}.  The relative astrometry and
photometry among the A, B, and C components derived from these data
was estimated by PSF fitting with IRAF/DAOPHOT, except for the 2008-9
epochs where the A and B components are within one diffraction ring
radius (e.g.~Fig.~\ref{fig:v773TauImage}).  For these epochs we used
the inner 0.1\arcsec~(the ``core'') of the C component as a PSF
template to model A component light, and estimated B parameters based
on residuals to this modeling.  The plate scales for NIRC2 and NACO
data were taken from \citet{Ghez2008} and \citet{Chauvin2010}
respectively.  Figure~\ref{fig:ABorbit} depicts both older (D2003) and
new (Table~\ref{tab:observations}) relative A-B astrometry.
Significant A-B orbital evolution since the D2003 summary is evident
in the new astrometry; we discuss modeling the A-B orbit in
\S~\ref{sec:orbit}.

\begin{figure}[t]
\epsscale{0.8}
\plotone{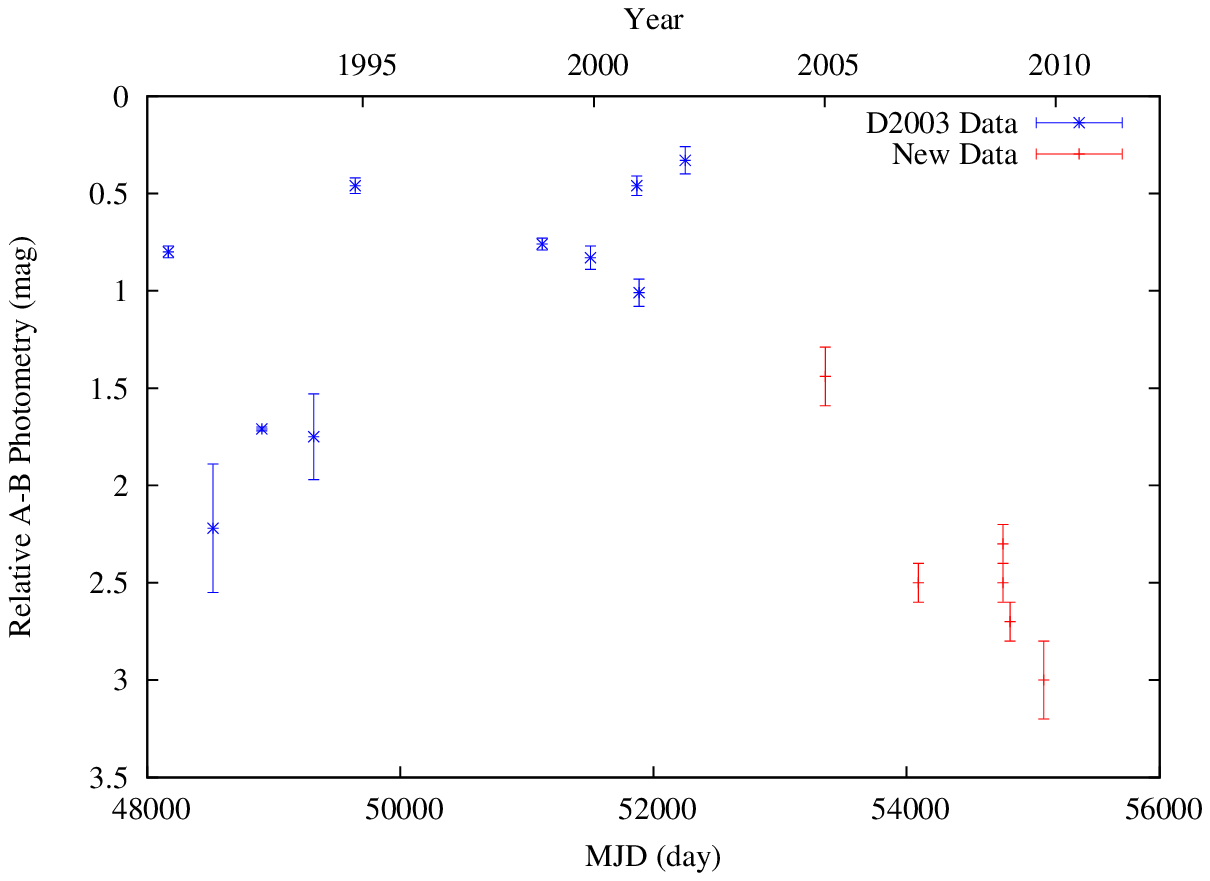}
\caption{Relative V773~Tau A-B $K$-band Photometry.  Here we show
  relative $K$-band photometry between the A and B components derived
  from resolved imaging.  Depicted in blue are older relative
  photometry data from D2003 Table 1; shown in red are new relative
  photometry data from this work (Table~\ref{tab:observations}).  Because
  the A component is thought to be photometrically stable (D2003),
  this variability is presumably in the B component.
\label{fig:relPhotometry}}
\end{figure}

Figure~\ref{fig:relPhotometry} depicts the relative photometry between
the V773 Tau A and B components for data with center-band wavelengths
within the near-IR $K$-band (2.0 -- 2.4 $\mu$m), including both older
data from D2003 (D2003 Table 1) and the new imaging presented here.
It is apparent that there is significant relative variability between
the two components; the data indicate relative $K$ variability of up
to 2.5 magnitudes -- nearly a magnitude more than had been reported in
D2003.  D2003 argued that the A subsystem is photometrically stable,
so presumably this large variability is due to the B component.  D2003
further argued that this variability resulted from circumstellar
material, an issue we will return to in \S~\ref{sec:discuss}.

\begin{deluxetable}{|cccc|ccc|ccc|}
\tablecolumns{10}
\tablewidth{0pc}
\tabletypesize{\small}
\rotate

\tablecaption{New V773~Tau Adaptive-Optics Imaging Summary
 \label{tab:observations} }

\tablehead{
\colhead{Epoch}  & \colhead{MJD}  & \colhead{Tel/Inst} &  \colhead{Filter} &
  \multicolumn{3}{c}{A-B} & \multicolumn{3}{c}{A-C} \\
  &  &  &  & 
 \colhead{$\rho$ (mas)} & \colhead{PA (deg)} & \colhead{$\Delta$m (mag)} & 
 \colhead{$\rho$ (mas)} & \colhead{PA (deg)} & \colhead{$\Delta$m (mag)} 
} 

\startdata


2004 Dec 19 & 53358  &  KeckII/NIRC2 & $K$
  & 112.8 $\pm$ 1.9  &  109.2 $\pm$ 1.2  &  1.44 $\pm$ 0.15
  & 249 $\pm$ 4      &  151.7 $\pm$ 1.4  &  2.2  $\pm$ 0.7  \\

2006 Dec 24 & 54093  &  VLT UT4/NACO & $K_S$
  &  95.3 $\pm$ 1.3  &  121.9 $\pm$ 1.3  &  2.5  $\pm$ 0.1
  & 254.5 $\pm$ 1.3  &  152.4 $\pm$ 1.0  &  2.08 $\pm$ 0.02  \\

2008 Oct 22 & 54761  &  KeckII/NIRC2 & $K_C$
  &  55.0 $\pm$ 5.0  &  137.9 $\pm$ 5.4  &  2.3  $\pm$ 0.1
  & 250.0 $\pm$ 1.8  &  155.3 $\pm$ 0.3  &  1.95 $\pm$ 0.03  \\

2008 Oct 22 & 54761  &  KeckII/NIRC2 & $K'$
  &  55.6 $\pm$ 1.8  &  143.7 $\pm$ 1.4  &  2.5  $\pm$ 0.1
  & 249.1 $\pm$ 1.0  &  155.2 $\pm$ 0.2  &  2.10 $\pm$ 0.02  \\

2008 Dec 18 & 54818  &  KeckII/NIRC2 & Br-$\gamma$
  &  49.0 $\pm$ 3.0  &  148.2 $\pm$ 4.5  &  2.7  $\pm$ 0.1
  & 250.5 $\pm$ 1.0  &  155.6 $\pm$ 0.3  &  2.41 $\pm$ 0.01  \\

2009 Sep 09 & 55083  &  KeckII/NIRC2 & $H_C$
  &  48.0 $\pm$ 2.0  &  175.6 $\pm$ 6.0  &  3.3  $\pm$ 0.2
  & 244.8 $\pm$ 1.2  &  156.3 $\pm$ 0.3  &  3.10 $\pm$ 0.03  \\

2009 Sep 09 & 55083  &  KeckII/NIRC2 & $K'$
  &  45.0 $\pm$ 2.0  &  181.3 $\pm$ 2.5  &  3.0  $\pm$ 0.2
  & 246.0 $\pm$ 1.0  &  156.9 $\pm$ 0.2  &  2.32 $\pm$ 0.03  \\

2009 Nov 29 & 55164  &  KeckII/NIRC2 & $K'$
  &  49.0 $\pm$ 4.0  &  185.8 $\pm$ 3.5  &  3.1  $\pm$ 0.2
  & 244.0 $\pm$ 2.0  &  156.5 $\pm$ 0.3  &  2.27 $\pm$ 0.04  \\

\enddata

\end{deluxetable}

\subsection{Radial Velocities}
\label{sec:A-RV}

\citet{Welty1995} established V773~Tau~A as an SB2 in spatially
unresolved optical spectroscopy, and \citet{Welty1995} and B2007
estimated the A orbit from RV measurements of both components.
Remarkably, no signs of other V773~Tau components have been found in
such optical spectroscopy -- a point we will return to in
\S~\ref{sec:discuss}.  Since B2007 we have continued to monitor A in
high-resolution optical spectroscopy, and have obtained eight
additional RV observations of both A components from the 1.5m
telescope on the Fred L. Whipple Observatory (FLWO) on Mt. Hopkins, AZ
(see discussion in B2007).  Further, the entire RV dataset has been
re-reduced for this analysis, now including both corrections for the
systematic effects of finite spectrometer passband \citep[a discussion
  of these corrections is given in][]{Torres1997}, and individual RV
uncertainties estimated from the corresponding spectra SNR (as opposed
to ensemble statistics as was done in B2007).  The resulting
refinement of the A-subsystem orbit model with this revised RV set is
discussed in T2011.


Additionally, for the purposes of A-B orbit analysis here the
double-lined A RV measurements also serve to probe the A barycenter
kinematics, complementing the A-B astrometry.  Straightforwardly, the
A-barycenter RV is derivable from A-component RV and the component
mass ratio ($q_A \equiv M_{Ab} / M_{Aa} = K_{Aa} / K_{Ab}$) as:
\begin{equation}
RV_A = \frac{M_{Aa} RV_{Aa} + M_{Ab} RV_{Ab} }{M_{Aa} + M_{Ab}} 
     = \frac{RV_{Aa} + q_A RV_{Ab} }{1 + q_A} 
\label{eq:barycenterRV}
\end{equation}
Equation~\ref{eq:barycenterRV} codifies the coupling of the A- and A-B
subsystem orbits in the A-component RV observables.  We return to the
joint modeling of the A and A-B orbits with these observables in
\S~\ref{sec:orbit}.


Table~\ref{tab:RVdata} gives a full listing of all A-subsystem
component RV and derived A-barycenter RV for $q_A$ = 0.831 $\pm$ 0.031
(see \S~\ref{sec:orbit}); these A-barycenter RV and A-B orbit model
are depicted in Figure~\ref{fig:ABorbit}.  (Note that the A-component
velocities in Table~\ref{tab:RVdata} are presented {\em without}
correction for A-subsystem motion.  This is different from the
presentation in B2007, where a model of the A-barycenter motion was
removed from the velocities reported in B2007 Table 3.)  The derived
A-barycenter velocities show roughly 5 km s$^{-1}$ of scatter on short
timescales; this is consistent with the single-measurement precision
of the V773 Tau Ab RV (see B2007 Table 3 and here Table 2).  But the
derived A-barycenter velocities also show clear and variable
acceleration over time, and a 13 km s$^{-1}$ offset between the two RV
data segments separated by $\sim$ 5000 days.  It is evident that these
derived A-barycenter velocities contain useful kinematic information
for modeling the A-B orbit.

\begin{table}
\dummytable\label{tab:RVdata}
\end{table}

\subsection{Spatially-Resolved Near-Infrared Spectroscopy}

In order to determine the spectral type of V773 Tau B, and in an
attempt to resolve the A component into a near-infrared double-lined
spectroscopic binary for the first time, we observed V773 Tau using
the near-infrared cross-dispersed spectrograph NIRSPEC
\citep{McLean2000} behind the Keck II AO system in two epochs (2003
Dec 11 and 2006 Dec 16, Table~\ref{tab:observations}).  In both cases,
we obtained $K$ band spectra aligning the slit with the A-B subsystem.
We obtained six 150s-integration with the 0\farcs013 slit in 2003, and
four 300s frames with the 0\farcs041 slit in 2006 in AB dither
patterns along the length of the slit. In both epochs, we observed an
early A-type star (\objectname[HD~27962]{HD~27962/HR~1389} in 2003,
\objectname[HD~34203]{HD~34203/HR~1718} in 2006) immediately following
V773 Tau to estimate the telluric transmission.  The resolution of the
spectra from 2003 is R$\sim$40000 and from 2006 is R$\sim$25000.

The basic data reduction was performed with REDSPEC, a software
package designed for
NIRSPEC\footnote{http://www2.keck.hawaii.edu/inst/nirspec/redspec/index.html}.
Object frames are reduced by subtracting opposing nods to remove sky
and dark backgrounds, dividing by a flat field, and correcting for bad
pixels.  Individual spectral orders are spatially rectified by fitting
the trace of the A0 calibrators with third order polynomials which are
then applied to target images.  The wavelength solution is determined
using the etalon lamps that are part of the NIRSPEC lamp suite
\citep{Figer2003}.  The absolute value of the etalon lines are
calibrated using the telluric features in our A0 calibrators
\citep{Konopacky2010}.  The wavelength solution is modeled with a
second-order polynomial.

Extraction of the A and B components from the reduced and rectified
frames is a challenge as cross-contamination is possible.  This is
particularly true of the 2006 epoch, where the separation was less
than 0\farcs1 and the A/B flux ratio had increased from previous
epochs (e.g.~Table~\ref{tab:observations},
Figure~\ref{fig:relPhotometry}).  In both 2003 and 2006, we opted to
extract the spectra by fitting a Gaussian to the trace of one visual
component and subtracting the fit result from the frame to leave only
the other component.  In 2003, we allowed the FWHM of the Gaussian to
vary with wavelength. In 2006, the two components are too closely
blended to fit for the FWHM; instead, we fixed the FWHM to the value
found for the telluric calibrator. The remaining trace in the frame
was then extracted by multiplying the flux by a normalized Gaussian
fit by the same method \citep{Konopacky2010}.  We then removed the
telluric features from the spectra by dividing by the extracted A0
calibrator spectra.  In order to use the A0 star to correct the order
containing Br$\gamma$, we fit the Br$\gamma$ line in the A0 stars with
a Lorentzian profile, which we subtracted out of the spectrum before
dividing.

\begin{figure}[hp]
\epsscale{1.0}
\plotone{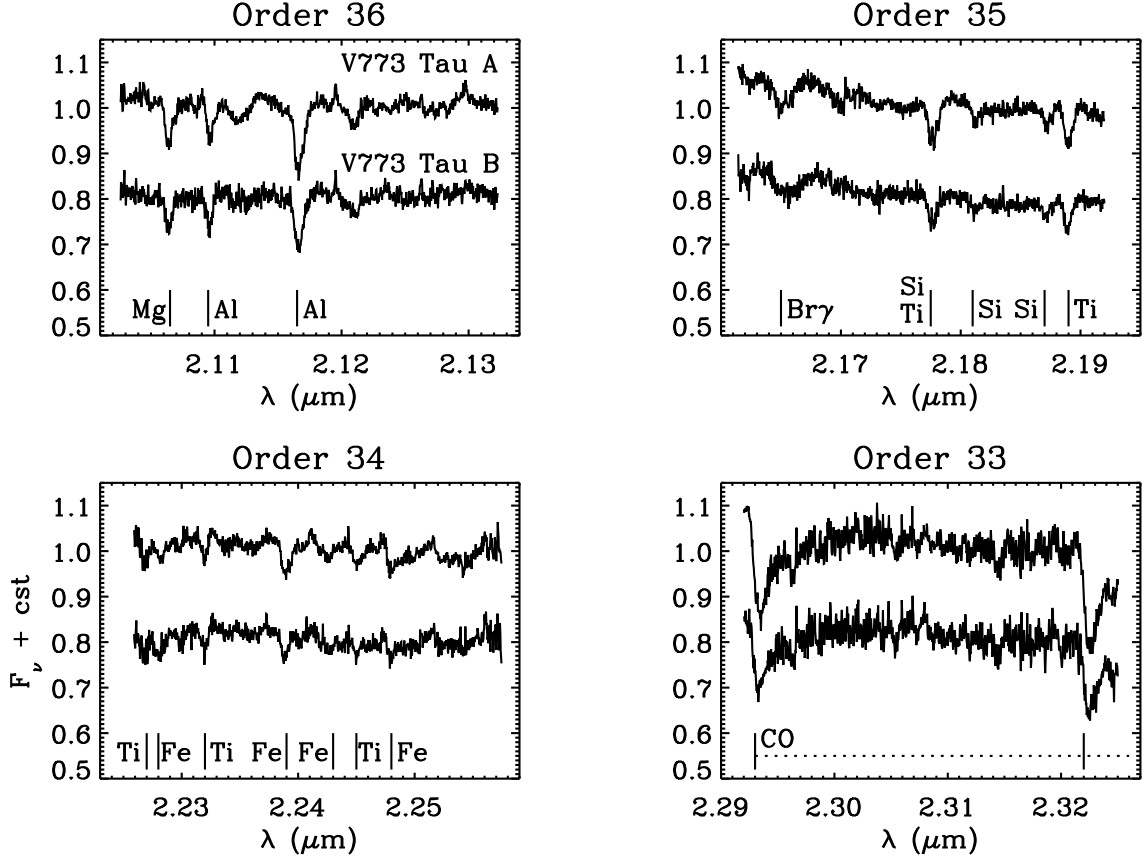}
\caption{Spectra of V773~Tau A and B obtained with NIRSPEC in Dec
  2003, when the pair had a relative astrometry and a flux ratio such
  that contamination from one component on the other is
  negligible. Each panel represent a separate order of the
  cross-dispersed spectra. In each panel, the spectra for both
  components were normalized by their median flux across the order;
  the spectrum for V773 Tau B was further displaced vertically by 0.2
  for clarity. Note that the continuum surrounding the Br$\gamma$ line
  (short wavelength end of the top right panel) has residual structure
  due to an imperfect interpolation of the photospheric line in the
  spectrum of the telluric standard.
  \label{fig:Spectra}}
\end{figure}

There are four useful orders for which spectra were extracted,
covering the approximate ranges 2.10--2.13 $\mu$m, 2.16--2.19 $\mu$m,
2.22--2.26 $\mu$m and 2.29--2.32 $\mu$m.  Samples of these spectral
orders are given in Figure~\ref{fig:Spectra}.

\section{Analysis}
\label{sec:analysis}

\subsection{A-B Orbit Modeling}
\label{sec:orbit}
At the time of D2003 considerable orbital evolution was apparent in
resolved imaging (and resulting relative astrometry) of the V773 Tau
A-B pair \citep[e.g.][D2003;
  Figure~\ref{fig:ABorbit}]{Tamazian2002,Woitas2003}, but it had not
yet completed even half a full orbit since the 1990 discovery of B
\citep{Ghez1993}.  Both \citet[][herein T2002]{Tamazian2002} and D2003
made preliminary estimates of the A-B orbit based on early astrometry,
but with the limited phase coverage of the data these two analyses
came to significantly different conclusions about the A-B orbit.

In the present work we have augmented the D2003 astrometry dataset
with the new astrometry presented above (Table~\ref{tab:observations},
plus one new astrometric point presented in \citet{Woitas2003}), and
the radial velocity observations presented above
(Table~\ref{tab:RVdata}).  Figure~\ref{fig:ABorbit} depicts expanded
relative astrometric and radial velocity data sets and our updated
orbital modeling (Table~\ref{tab:ABorbit}) for V773~Tau A-B.  In
particular the left panel depicts the available set of relative
astrometry on the A-B pair, and both the D2003 and our orbit model
visual trace.  Older astrometry summarized in D2003 (specifically
D2003 Table 1) is rendered in blue, while newer astrometry derived
from imaging reported here is rendered in red.
Figure~\ref{fig:ABorbit} right panel shows (derived) A subsystem
barycentric radial velocity (\S~\ref{sec:A-RV};
Table~\ref{tab:RVdata}) and the radial velocity trace from our model
orbit.  Figure~\ref{fig:ABorbit} right also shows a rendering of the
A-barycenter RV smoothed at the A-subsystem period (51.1d; B2007).
This rendering is provided to illustrate the A-barycenter RV content
relevant to the A-B orbit modeling, these smoothed data are not used
in the orbit modeling.

\begin{deluxetable}{l|cc|c}
\tabletypesize{\footnotesize}
\tablecolumns{4}
\tablewidth{0pc}

\tablecaption{Orbital Parameters for V773~Tau A-B \label{tab:ABorbit}}


\tablehead{
\colhead{Orbital}   & \colhead{T2002} & \colhead{D2003} & \colhead{Joint Solution} \\
\colhead{Parameter} &                 &                 & \colhead{(This Work)}
}

\startdata
Period (yr)         & 125.0 $\pm$ 6.0     & 46.0 $\pm$ 6.0      & 26.20  $\pm$ 1.1         \\
T$_{0}$ (yr)         & 1998.62 $\pm$ 3.0   & 1996.5 $\pm$ 0.8   & 2010.53 $\pm$ 1.0         \\
$e$                 & 0.643 $\pm$ 0.04    & 0.30 $\pm$ 0.10     & 0.099 $\pm$ 0.026        \\ 
K$_A$ (km s$^{-1}$)  &                     &                     &  6.50 $\pm$ 0.50           \\
$\gamma$ (km s$^{-1}$)  &                  &                     & 16.30 $\pm$ 0.51         \\
$\omega_{A}$ (deg)   & 299.4 $\pm$ 10.0    & 81 $\pm$ 10         & 94  $\pm$  17             \\
$\Omega$ (deg)      & 101.1 $\pm$ 7       & 288 $\pm$ 1         & 288.2 $\pm$ 1.0           \\ 
$i$ (deg)           & 63.8  $\pm$ 5       & 66 $\pm$ 3          & 71.48 $\pm$ 0.78          \\ 
$a$ (mas)           & 249   $\pm$ 15     & 140 $\pm$ 10         & 115.5 $\pm$ 3.4           \\ 
\hline
\hline
\enddata

\tablecomments{Summarized here are the orbital parameters for the A-B
  subsystem as estimated by T2002, D2003, and present results from our
  joint modeling of the V773~Tau A and A-B orbits.  $\omega_{A}$ is
  the argument of periastron for the A subsystem, and $\Omega$ is
  quoted in a position angle convention.  $\Omega$ and $\omega_{A}$
  from T2002 appear particularly discrepant from the other work
  summarized here.  We present these parameters as they are listed in
  T2002 Table~2, but the general agreement of the A-B orbit
  orientation depicted in their Figure~1 and our
  Figure~\ref{fig:ABorbit} leads us to speculate that $\Omega$ and
  $\omega_{A}$ may have been reversed as they are listed in T2002 --
  bringing them into much better agreement with other results.}

\end{deluxetable}

Table~\ref{tab:ABorbit} lists the parameters for the A-B orbit as
estimated here and in the earlier work from T2002 and D2003.  Using
input relative astrometry (e.g.~Table~\ref{tab:observations}) and
A-barycenter RV (Table~\ref{tab:RVdata}) we made initial estimates of
the A-B orbit using both Marquardt-Levenberg least-squares and
Bayesian modeling techniques, and found good agreement in the results
from these two methods \citep[see orbit modeling descriptions
  in][]{Boden2000,Torres2002,Boden2005}.  Our final refinement of the
A-B orbit model was made by joint modeling of both the A and A-B
subsystem orbits simultaneously, coupling the A-component RV data to
the two subsystem orbits through Eq.~\ref{eq:barycenterRV}, and adding
A-subsystem relative astrometry data from the Keck Interferometer
(B2007) and VLBA (B2007, T2011) sources.  In practice the joint
modeling was accomplished by iterative application of
Marquardt-Levenberg optimization for the A and A-B models in turn,
coordinating the A and A-B RV observables through
Eq.~\ref{eq:barycenterRV} between iterations.  This process converged
in only a few (four) iterations, and resulted in stable solutions for
both orbits.  The A-subsystem orbit solution is described in the
companion paper (T2011) and is found to be in good statistical
agreement with the earlier estimate from B2007.  It is both notable
and relevant to the A-B results that the new RV reduction and joint
A/A-B modeling yields a subtly different value for $q_A$ (0.831 $\pm$
0.031) than found in B2007 (0.865 $\pm$ 0.032).  Additional details on
the final A orbit model can be found in T2011.


Focusing here on the A-B portion of the joint solution, we find
significant differences between our A-B orbit modeling results and
those published previously by T2002 and D2003; e.g.~see the visual
orbits rendered in Figure~\ref{fig:ABorbit}.  Prima facie these
differences are due to the greatly expanded data set available here to
estimate the A-B orbit.  Our modeling is supported both by expanded
phase coverage in the relative astrometry
(e.g.~Figure~\ref{fig:ABorbit}), and the addition of complementary
A-barycenter RV data, neither of which was available in earlier work.
At present epoch the astrometric dataset covers roughly 73\% of the
estimated orbit period, compared to only 46\% for data available to
D2003.  Further, the new astrometric data is seen to be particularly
important in revealing the orbit's character: Figure~\ref{fig:ABorbit}
shows that the A-B apparent orbit passed maximum elongation around the
time of D2003, and since that time has exhibited remarkably rapid
position-angle evolution as the system approaches periastron at
roughly the present epoch (Table~\ref{tab:ABorbit}).  These
differences are most pronounced in the A-B orbit period -- our present
period estimate (26.2 yr) is roughly half of the value estimated by
D2003, and a factor of five smaller than estimated by T2002.  Finally
the addition of the A-subsystem RV is seen to strongly support the
updated A-B orbit modeling.

\subsection{V773~Tau~B Dynamical Mass}
\label{sec:dynamicalMass}
Typically a relative astrometric orbit such as derived here requires
the support of radial velocities for both stars \citep[e.g.][B2007,
  and references therein]{Torres2002} in order to yield unambiguous
component dynamical masses.  However V773~Tau~A-B is unique in that it
is supported by the A subsystem binarity, and the analyses of the A
subsystem by B2007 and T2011 
-- providing both A
subsystem total mass and distance.  When combined with these two items
the orbit model in Table~\ref{tab:ABorbit} allows an unambiguous
estimate for the B component dynamical mass.

\begin{deluxetable}{lcc}
\tabletypesize{\footnotesize}
\tablecolumns{3}
\tablewidth{0pc}

\tablecaption{Physical Parameters for V773~Tau A-B \label{tab:ABphysics}}

\tablehead{
\colhead{Parameter}  & \colhead{Value}  & \colhead{Note}
}

\startdata

System Distance (pc)        &  132.8 $\pm$ 2.4     &  T2011 \\

A-subsystem Mass (M$_\sun$)     &  2.91 $\pm$ 0.20  &  B2007; T2011 \\

A-subsystem Luminosity (L$\sun$) & 3.93 $\pm$ 0.38 &  B2007  \\

A-B Semi-Major Axis (AU)    &  15.35 $\pm$ 0.45    &   \\

A-B System Mass (M$_\sun$)   &  5.27 $\pm$ 0.65     &   \\

B Mass (M$_\sun$)            &  2.35 $\pm$ 0.67     &  \S~\ref{sec:dynamicalMass} \\

B Luminosity (L$_\sun$)      &  2.6 $\pm$ 0.6       &  \S~\ref{sec:discuss} \\

\enddata

\tablecomments{Summarized here are the physical parameters for the A-B
  subsystem as given by previous results and derived here.  Note we
  use the preferred composite distance estimate from T2011 as the
  basis for these computations, but other variations are possible --
  please refer to comments in the text for more detail.}


\end{deluxetable}

Table~\ref{tab:ABphysics} summarizes the computed physical properties
for the A-B components.  In particular the new quantities provided by
the A-B orbit are the A-B physical semi-major axis and B dynamical
mass.  Adopting the T2011-favored composite system distance of 132.8
$\pm$ 2.4 pc, the resulting A-B semi-major axis is 15.35 $\pm$ 0.45
AU, and the B dynamical mass is 2.35 $\pm$ 0.67 M$_\sun$.  This large
value for the B component mass is completely unexpected (e.g.~see
discussion in D2003), and makes B's mass comparable to that for A.
However note that the T2011-estimated A subsystem mass corresponds to
an A-subsystem orbital distance of 135.7 $\pm$ 3.2 pc.  Using the
A-subsystem orbital distance to interpret the A-B orbit presented here
(thereby placing the A and B dynamical masses on a consistent distance
basis), the resulting SMA and B mass are 15.67 $\pm$ 0.46 AU and 2.69
$\pm$ 0.67 respectively.  While we recommend the former values based
on the T2011 composite distance estimate as probably more accurate,
the latter values represent viable alternate (and statistically
consistent) estimates based on present data.

\begin{figure}[th]
\epsscale{0.65}
\plotone{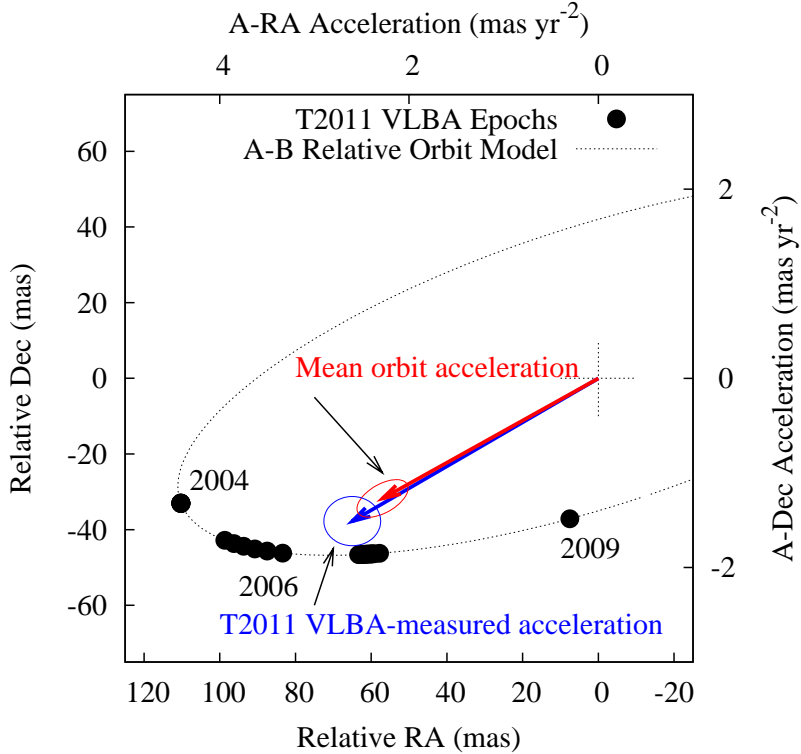}
\caption{V773 Tau A Acceleration Estimates.  The orbital acceleration
  on A produced by B is measured by the VLBA astrometry reported by
  T2011.  The relative A-B position is shown here for the 27 T2011
  VLBA measurement epochs; we can use our A-B physical orbit model and
  the component (A: B2007, T2011, and B: this work) dynamical mass
  values to estimate the mean acceleration in the T2011 data.  The
  mean A acceleration estimates from T2011 (blue) and our orbit model
  (red) are shown along with their 1-$\sigma$ uncertainty ellipses.
  The two acceleration estimates are seen to be in excellent
  (i.e.~well within 1-$\sigma$) agreement, providing an independent
  confirmation of our orbit model and B dynamical mass estimate.
\label{fig:Aaccel}  }
\end{figure}

The companion VLBA study of T2011 offers the unique opportunity to
cross-check our B dynamical mass estimate through independent means.
In modeling the VLBA data on V773~Tau~A, T2011 
find it necessary to account for the A-B orbital motion as an
effective mean acceleration over the time interval of their data.
This situation is depicted in Figure~\ref{fig:Aaccel}.  At each of the
epochs of the T2011 VLBA observations the B mass accelerates A toward
B's position.  With our A-B orbit model (Table~\ref{tab:ABorbit}), B
dynamical mass, and system distance (Table~\ref{tab:ABphysics}) we can
estimate the instantaneous A acceleration, and the resulting average
acceleration over the 27 VLBA epochs (T2011 Table 1).  When we do, we
estimate the average sky-projected acceleration to be $a_\alpha \cos
\delta$ = 2.28 $\pm$ 0.65 mas yr$^{-2}$, $a_\delta$ = -1.27 $\pm$ 0.36
mas yr$^{-2}$ As shown in Figure~\ref{fig:Aaccel}, this value is
well-within 1-$\sigma$ agreement with the VLBA-measured mean
acceleration ($a_\alpha \cos \delta$ = 2.60 $\pm$ 0.60 mas yr$^{-2}$,
$a_\delta$ = -1.51 $\pm$ 0.52 mas yr$^{-2}$; vector and uncertainty
ellipse rendered in blue) from T2011.  The consistency of these two
independent acceleration estimates imply that our A-B orbit model and
B dynamical mass are reliable at their stated uncertainties.

\subsection{Near-Infrared Spectroscopic Results for V773 Tau B}
\label{sec:specAnalysis}

The extracted NIRSPEC spectra for V773~Tau A and B are shown in
Figure~\ref{fig:Spectra}.  At first glance, the $K$-band spectra of
V773 Tau A and B appear very similar, suggesting that the B component
has a K-type spectral type, similar to A.  Unfortunately, our
cross-dispersed spectra do not include the strongest photospheric
features of late-type stars (Na doublet at 2.20 $\mu$m, Ca triplet at
2.26 $\mu$m), so we have to rely on somewhat weaker features to
estimate the spectra type of V773 Tau B. The strongest features in our
spectra are an Mg doublet and two Al lines in order 36, H Br$\gamma$,
Si, Ti and Fe lines in order 35, weak Fe and Ti lines in order 34, and
the CO (2--0) forest and the first two bandheads in order 33.  The
various line ratios are very similar for components A and B, although
all lines appear weaker in B than in A. This may be due to
contamination by emission from hot circumstellar dust (veiling) or by
the presence of a hotter (hence almost featureless) unresolved
component.

We compared our spectra to several published libraries of $K$ band
spectra for field stars
\citep{Kleinman1986,Wallace1996,Ivanov2004,Rayner2009} as well as to
the Gemini/GNIRS spectral
library\footnote{http://www.gemini.edu/sciops/instruments/nearir-resources/?q=node/10167}
to estimate the spectral type of both components. For the A component,
we infer a K1--K3 spectral type in good agreement with the system's
consensus spectral typing, while for the B component, we find a K2--K5
spectral type. The slightly later type for B is driven by 1) the
slightly weaker Si line at 2.187 $\mu$m (with respect to the nearby
Ti-Fe blend at 2.189 $\mu$m), and 2) the slightly more marked
``shoulder'' just longward of the first CO bandhead around 2.295
$\mu$m. Still, considering the limited sampling in spectral types of
the libraries we used and the possibility of veiling in the B
component, we acknowledge that both components may well have the same
spectral type and that both estimates could be systematically offset.

\begin{figure}[t]
\epsscale{1.1}
\plottwo{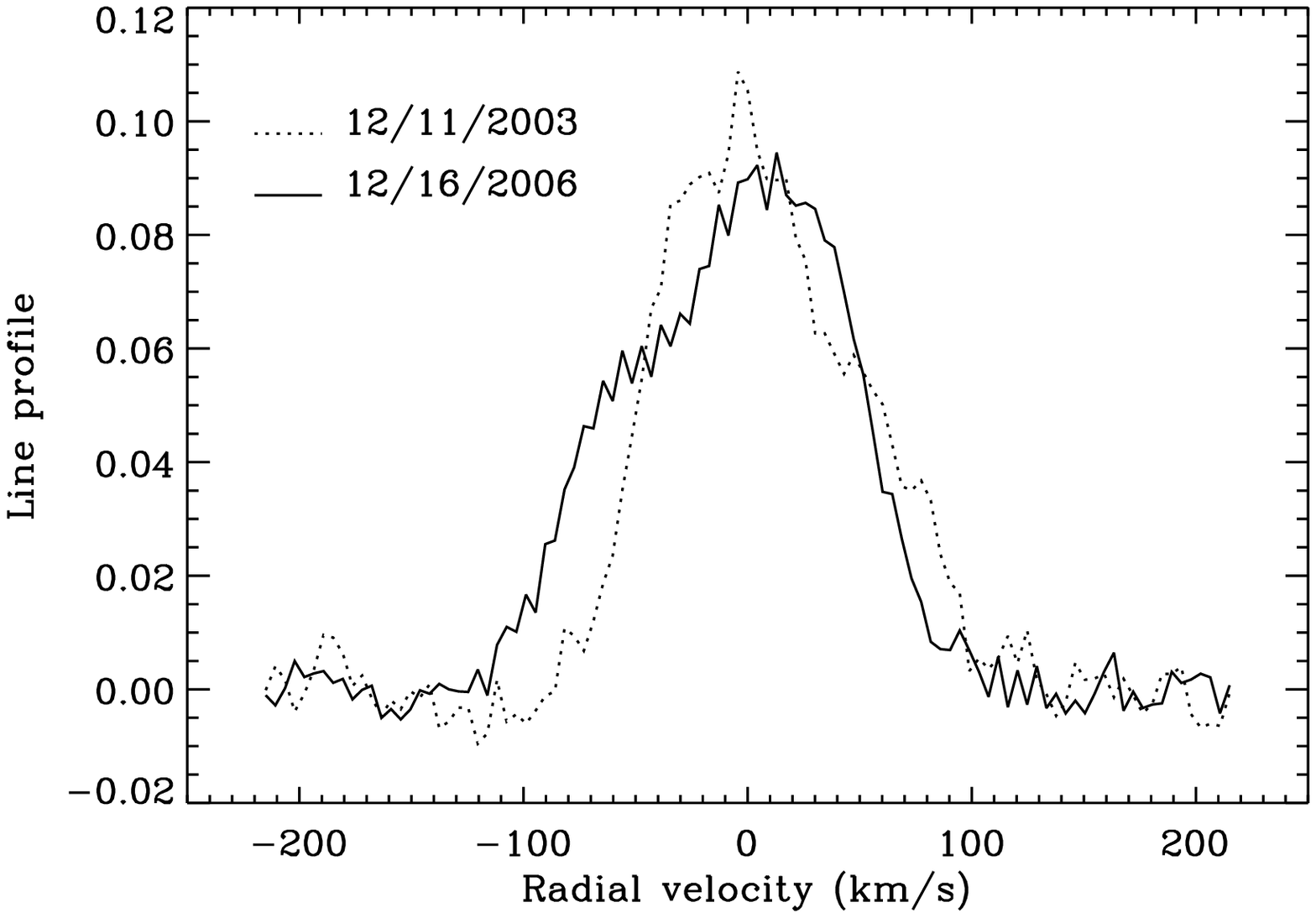}{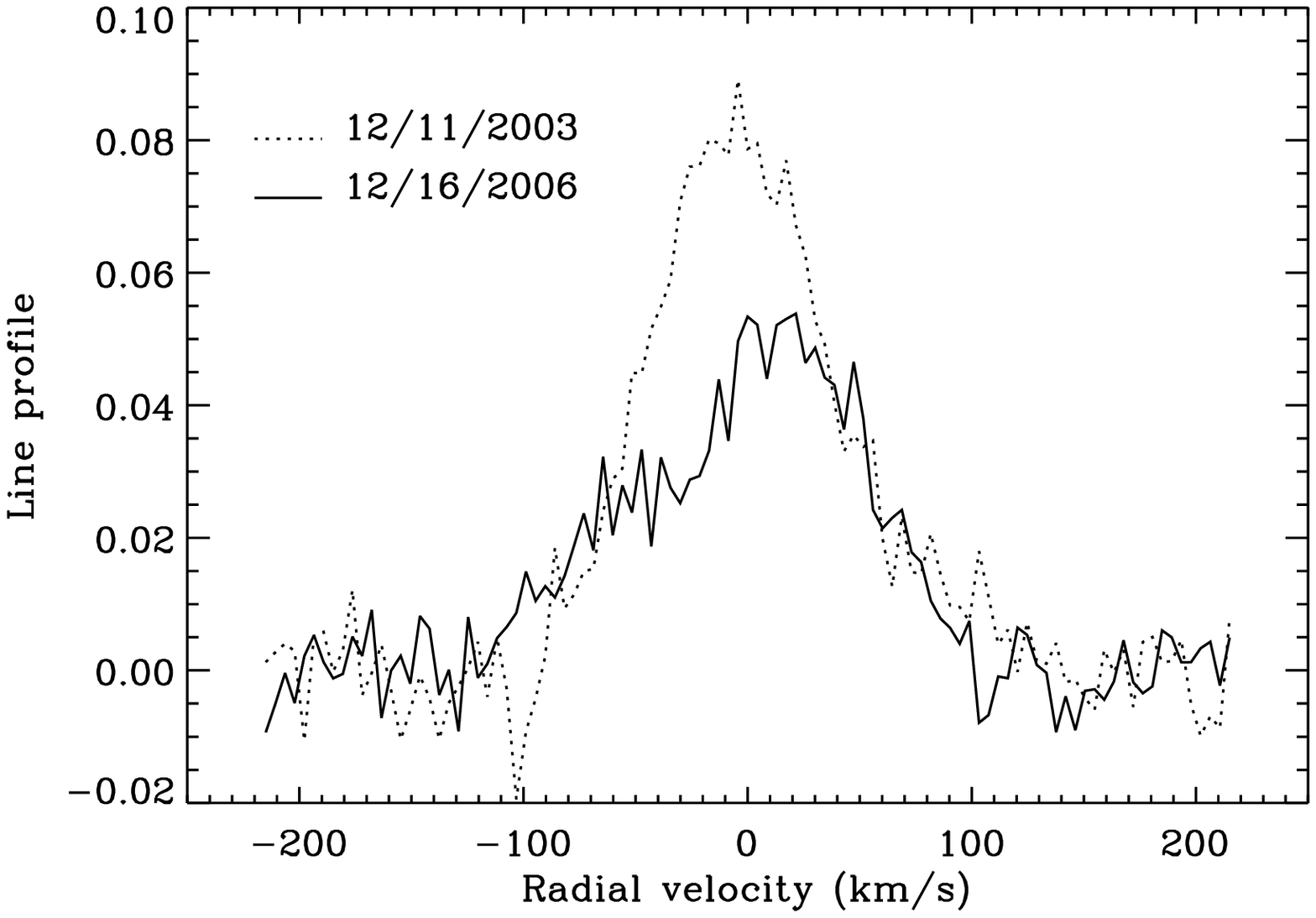}
\caption{Average line profile for V773 Tau A (left) and V773 Tau B
  (right) based on the four most prominent isolated lines in our
  NIRSPEC spectra.
  \label{fig:LineProfile}}
\end{figure}

Inspection of the V773~Tau~A spectra revealed that the profile of
the strongest photospheric lines were asymmetric in both epochs, with
a red ``shoulder'' in 2003 and a ``blue'' one in 2006.  To illustrate
this point we extracted the profiles of the four strongest isolated
lines in our spectra (Mg 2.107$\mu$m, Al 2.110$\mu$m, Al 2.117$\mu$m
and Ti 2.189$\mu$m -- the line split of the Ti/Si 2.179$\mu$m doublet
is just large enough to prevent us from using this feature in the
average profiles) and computed the average profile using the relative
strength of the line as a weight.  The resulting line profiles
(Figure~\ref{fig:LineProfile} left panel) indicate that we have
resolved V773~Tau~A as a double-lined spectroscopic binary in the
near-infrared, despite the relative large rotational broadening
(B2007).  The Ab component is detected offset from Aa at about +60
km/s in 2003 and at about $-$75 km/s.  These velocity offsets are in
good agreement with the predictions from A-subsystem orbit modeling
(+54.4 km/s and $-$70.9 km/s in 2003 and 2006 respectively; B2007).

We further generated similar average line profiles for V773~Tau~B in
both epochs.  We find the average line profile to be symmetric in 2003
but much broader and most likely double-lined in 2006 (with a relative
radial velocity of about $-$80 km/s for component Bb relative to Ba). If
confirmed in future observations of the system, this is the first
direct evidence supporting the inference that V773~Tau~B is itself an
unresolved binary.

\section{Discussion}
\label{sec:discuss}

The most notable result from this analysis is the large value for the
V773~Tau~B dynamical mass, 2.35 $\pm$ 0.67 M$_\sun$.  This B-mass
value is implied by the A-B orbit modeling described here, and
independently supported by the companion VLBA study of T2011.  The B
dynamical mass estimate makes B comparable to, but slightly
less massive than the A subsystem.  Further, this B mass, along with
the estimated luminosity (see below) and signs of double-lined
features in our near-IR spectra (\S~\ref{sec:specAnalysis}) suggests
that like A, B is also a multiple stellar system.  Before we discuss
the implications of these results we need to consider the luminosity
of V773~Tau~B.

In modeling the V773~Tau~B SED, D2003 found a range of possible
solutions as a function of possible line-of-sight extinction.  In
particular:
\begin{quote}
``With an additional 1 mag of extinction at $V$ toward V773 Tau [B], we
  find that its SED is well fitted by a K7 dwarf with a luminosity of
  2.3 $\pm$ 0.3 L$_\sun$; a significant excess at $L$' remains present
  although the $H$- and $K$-band fluxes are then consistent with
  photospheric levels.''
\end{quote}

\begin{figure}[t]
\includegraphics[angle=-90,width=14.5cm]{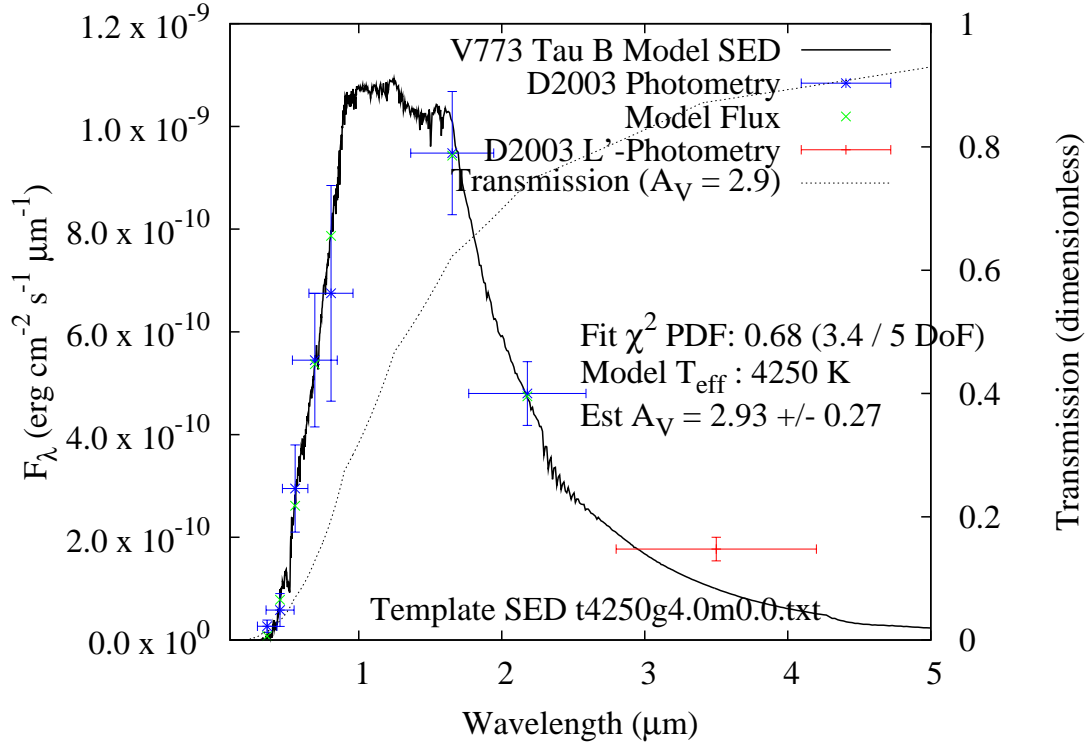}
\caption{Sample V773 Tau B Spectral Energy Distribution Model.  A
  sample SED model is shown using a 4250 K Kurucz-Lejunne SED template
  \citep{Lejeune1997} with significant line-of-sight extinction ($A_V$
  = 2.9 mag).  The implied B luminosity is 2.6 $\pm$ 0.5 L$_\sun$.  As
  first reported in D2003, the $L$' flux point (shown in [red])
  indicates there is a marginally significant excess compared with the
  estimated photospheric level.
  \label{fig:v773TauBSED}}
\end{figure}

Here we model the V773~Tau~B spectral energy distribution (SED)
informed by the mid-K spectral type determination from
\S~\ref{sec:specAnalysis}.  Resolved B photometry
\citep[][D2003]{White2001} are well-fit by a range of mid-K SED
templates (T$_{\rm eff} \sim$ 4100 -- 4300 K) from
\citet{Pickles1998}, \citet{Lejeune1997}, and PHEONIX
\citep[e.g.][]{Allard2000}, attenuated by considerable line-of-sight
extinction (e.g.~$A_{\rm V} \sim$ 3 mag).  To facilitate comparison
with B2007 results on V773~Tau~A, Figure~\ref{fig:v773TauBSED}
presents a model derived from the same template family
(solar-abundance Kurucz-Lejeune template at T$_{\rm eff}$ = 4250 K and
$\log$ g = 4.0).  This model clearly fits the available photometry well
($\chi^2$/DOF = 0.68), and results in a B-luminosity estimate of 2.6
$\pm$ 0.5 L$_\sun$ with $A_V$ of 2.9 $\pm$ 0.3 mag.  This extinction is
roughly one magnitude higher in $A_V$ than reported toward A ($A_V$ =
1.8 $\pm$ 0.2 mag) in B2007; taking the A extinction value as
interstellar this suggests a circum-B source for the $\sim$ one
magnitude of additional extinction found here.  Like D2003, our SED
modeling also finds a marginally significant $L$' excess, further
reinforcing the hypothesis that there is circum-B material
reprocessing radiation from B.  However, the interpretation of this B
SED modeling is complicated by the fact that B is seen to be variable
(Figure~\ref{fig:relPhotometry}), and the input data
\citep[][D2003]{White2001} are not contemporaneous.

The combination of our B-component dynamical mass (2.35 M$_\sun$) and
luminosity estimate (2.6 L$_\sun$) suggests a multiple system
interpretation.  For instance, \citet{DAntona1997} models predict a
luminosity of 17 L$_\sun$ and T$_{\rm eff}$ $\sim$ 6200 K for a 3 Myr
(B2007) single star at 2.35 M$_\sun$.  (Note that a 1-$\sigma$
excursion downward in the mass estimate -- 1.7 Msun -- would predict
2.8 L$_\sun$ and 5200 K from these same models.)  Such a B-luminosity
and temperature would seem to be impossible based on determinations
presented here.  However assuming the simplest multiple configuration
for B as a binary arrangement with a pair of 1.175 M$_\sun$ stars, the
\citet{DAntona1997} models predict a pair of such stars to have a
total luminosity of 2.6 L$_\sun$, and T$_{\rm eff}$ $\sim$ 4800 K, in
better agreement with the findings here.  While mass arguments alone
would not exclude a single star interpretation, combined with
double-lined indications in the near-IR spectroscopic analysis
(\S~\ref{sec:specAnalysis}), the circumstantial case that V773~Tau~B
is a multiple stellar system seems strong.  However, experience
(e.g. T2002, D2003) has shown that one must consider the present A-B
orbit model as preliminary pending additional phase coverage, and the
B-mass estimate is still significantly uncertain ($\sim$ 29\%
1-$\sigma$).

Photometric variability in V773 Tau has typically been attributed to
the B component (D2003; Figure~\ref{fig:relPhotometry}), and
interpreted as changing line-of-sight extinction from circum-B
material.  This interpretation is at least complicated both by
potential B multiplicity, and by the physical size of the A-B orbit
(15.3 AU, Table~\ref{tab:ABphysics}).  Dynamical studies
\citep[e.g.][]{Artymowicz1994,Pichardo2005} indicate that stable
regions for diffuse material are offset from binaries by several times
the orbital semi-major axis.  Presumably the outer radius of any
stable circum-B orbit should be on the order of 15 / 3 $\sim$ 5 AU.
Similarly, the typical size of a putative B subsystem should be
several times smaller than this 5 AU scale -- probably $\leq$ 1 AU --
making the working model for B a short-period binary.  In this
picture, circumbinary material in the approximately 2--5\,AU range
would be responsible for both the near-infrared excess and photometric
variability of B.  The large amount of dust needed to explain both
extinction and variability (see below) raises the question whether it
could remain dynamically stable in the region between B as a putative
short-period binary and the larger A-B orbit, and whether such a
reservoir of material would show other observable signatures such as
detectable thermal IR and millimeter flux.  As a system V773~Tau shows
a large mid-IR flux above photspheric levels \citep[e.g.][D2003,
  Spitzer Taurus Legacy program, Padgett et al.~2011 in
  prep]{Prusti1992}, and \citet{Andrews2005} lists significant flux at
850 $\mu$m and 1.3 mm.  However the presence of the V773~Tau~C
component \citep[D2003,][]{Woitas2003} significantly complicates the
interpretation of these flux measurements.  A spatially-resolved mm
study with a sensitive facility such as ALMA would clarify the
distribution of circumstellar material in the V773~Tau system.

If the putative B variability is due to changing line-of-sight
extinction, this degree of variability is extraordinary.  To account
for the amplitude of the $K$-variability shown in Figure~3, a
variation in $A_V$ of up to 25\,mag (assuming ISM-like dust opacity)
would be indicated.  Such large $A_V$ excursions seem discrepant with
the results of the SED modeling above.  A more plausible alternative
might be an optically-thick structure -- perhaps a disk that may or
may not be perturbed by complicated dynamics -- obscuring the
line-of-sight to B components, allowing only a small amount of
scattered light to reach the observer.  Variability on short timescale
could be related to one of the components of B temporarily coming into
direct view along its orbit, akin to the KH~15D system \citep[][and
  references therein]{Herbst2010}.  The current sampling of the light
curve for B is too sparse to decisvely test this hypothesis.  In
either case, this large attenuation could explain why no clear
detection of B is seen in extensive optical spectroscopic monitoring
\citep[B2007, Table~\ref{tab:RVdata}]{Welty1995}.


If the inference of B multiplicity and circum-B material is correct,
then the V773~Tau system seems remarkably similar to the
\objectname[HD 98800]{HD~98800} system, where circumbinary material is
found around one but not both binaries in the system
\citep{Prato2001}.  In HD~98800 the inner B and outer A-B orbit
significantly complicates the dynamics of B circumbinary material,
leading to an inference of disk truncation and warping
\citep{Boden2005,Akeson2007,Verrier2007,Pichardo2008}.  The apparent
increasing system complexity raises questions of stability in
V773~Tau.  While more information on the B and C components will be
necessary to address the specific stability of V773~Tau, it is clear
that such complex systems are a common outcome of the star formation
process, and can remain stable for Gyr
\citep[e.g.][]{Duquennoy1991,Tokovinin2008,Raghavan2010}.

\acknowledgements 

Some of the data presented here were obtained at the W.M.~Keck
Observatory, which is operated as a scientific partnership among the
California Institute of Technology, the University of California, the
University of Hawaii, and NASA.  The Observatory was made possible by
the generous financial support of the W.M.~Keck Foundation.  We
gratefully acknowledge personnel from the W.M.~Keck Observatory in
supporting observations of V773~Tau.  The authors wish to recognize
and acknowledge the very significant cultural role and reverence that
the summit of Mauna Kea has always had within the indigenous Hawaiian
community.  We are most fortunate to have the opportunity to conduct
observations from this mountain.

The authors gratefully acknowledge research support provided by the
National Science Foundation, the California Institute of Technology,
Harvard University, the University of California, DGAPA, UNAM, and
CONACyT, Mexico.  In particular: AFB acknowledges support from NSF
grant AST-0908822, GT acknowledges support from NSF grant AST-1007992,
RMT acknowledges support by the Deutsche Forschungsgemeinschaft (DFG)
through the Emmy Noether Research grant VL 61/3-1. LL acknowledges the
financial support of the Guggenheim Foundation and the von Humboldt
Stiftung.  Portions of this work was performed under the auspices of
the U.S. Department of Energy by Lawrence Livermore National
Laboratory under Contract DE-AC52-07NA27344.


\begin{thebibliography}{}

\bibitem[Allard et al.(2000)]{Allard2000}
Allard, F.~et al.~2000, \apj~539, 366.

\bibitem[Akeson et al.(2007)]{Akeson2007}
Akeson, R.~et al.~2007, \apj~670, 1240.

\bibitem[Andrews \& Williams(2005)]{Andrews2005}
Andrews, S.~\& Williams, J.~2005, \apj~631, 1134.


\bibitem[Artymowicz \& Lubow(1994)]{Artymowicz1994}
Artymowicz, P.~\&  Lubow, S.~1994 \apj~421, 651.



\bibitem[Boden et al.(2000)]{Boden2000}
Boden, A., et al.~2000, \apj~536, 880.

\bibitem[Boden et al.(2005)]{Boden2005}
Boden, A.~et al.~2005, \apj~635, 442.

\bibitem[Boden et al.(2007)]{Boden2007}
Boden, A.~et al.~2007, \apj~670, 1214 (B2007).


\bibitem[Chauvin et al.(2010)]{Chauvin2010}
Chauvin, G.~et al.~2010, \aap~509, 53.



\bibitem[D'Antona \& Mazzitelli(1997)]{DAntona1997}
D'Antona, F.~\& Mazzitelli, I.~1997, Mem.~S.A.It., 68, 807




\bibitem[Duch\^ene et al.(2003)]{Duchene2003}
Duch\^ene, G.~et al.~2003, \apj~592, 288 (D2003).

\bibitem[Duquennoy \& Mayor(1991)]{Duquennoy1991}
Duquennoy, A.~\& Mayor, M.~1991, \aap~248, 485.


\bibitem[Feigelson et al.(1994)]{Feigelson1994}
Feigelson, E.~et al.~1994, \apj~432, 373.


\bibitem[Figer et al.(2003)]{Figer2003}
Figer, D.~et al.~2003, \apj~599, 1193.


\bibitem[Ghez et al.(1993)]{Ghez1993}
Ghez, A., Neugebauer, G., \& Matthews, K.~1993, \aj~106, 2005.


\bibitem[Ghez et al.(1997)]{Ghez1997}
Ghez, A., White, R., \& Simon, M.~1997, \apj~490, 353.


\bibitem[Ghez et al.(2008)]{Ghez2008}
Ghez, A.~et al.~2008, \apj~689, 1044.

\bibitem[Hartigan et al.(1994)]{Hartigan1994}
Hartigan, P., Strom, K., \& Strom, S.~1994, \apj~427, 961.


\bibitem[Herbst et al.(2010)]{Herbst2010}
Herbst, W.~et al.~2010, \aj~140, 2025.


\bibitem[Ivanov et al.(2004)]{Ivanov2004}
Ivanov, V.~et al.~2004, \apjs~151, 387.





\bibitem[Kleinman \& Hall(1986)]{Kleinman1986}
Kleinman, S.~\& Hall, D.~1986, \apjs~62, 501.


\bibitem[Konopacky et al.(2010)]{Konopacky2010}
Konopacky, Q.~et al.~2010, \apj~711, 1087.


\bibitem[Kutner et al.(1986)]{Kutner1986}
Kutner, M., Rydgren, A., \& Vrba, F.~1986, \aj~92, 575.


\bibitem[Latham(1992)]{Latham1992}
 Latham, D.\ W. 1992, in IAU Coll.\ 135, Complementary Approaches to
Double and Multiple Star Research, ASP Conf.\ Ser.\ 32, eds.\ H.\ A.\
McAlister \& W.\ I.\ Hartkopf (San Francisco: ASP), 110


\bibitem[Latham et al.(2002)]{Latham:02}
 Latham, D.~W.~et al.~2002~\aj~124, 1144.



\bibitem[Leinert et al.(1993)]{Leinert1993}
Leinert, C.~et al.~1993, \aap~278, 129.


\bibitem[Lejeune et al.(1997)]{Lejeune1997}
Lejeune, T.~et al.~1997, \aaps~125, 229.

\bibitem[Lenzen et al.(2003)]{Lenzen2003}
Lenzen, R.~et al.~2003, Proc. SPIE~944, 4841.


\bibitem[Lestrade et al.(1999)]{Lestrade1999}
Lestrade, J-F, et al.~1999, \aap~344, 1014 (L1999).


\bibitem[Mason et al.(2001)]{Mason2001}
Mason, B.~et al.~2001, \aj~122, 3466.


\bibitem[Montalb\'an \& D'Antona(2006)]{Montalban2006}
Montalb\'an, J.~\& D'Antona, F.~2006, \mnras~370, 1823.

\bibitem[Mart\'in et al.(1994)]{Martin1994}
Mart\'in, E.~et al.~1994, \aap~282, 503.

\bibitem[Massi et al.(2002)]{Massi2002}
Massi, M., Menten, M., \& Neidhofer, J.~2002, \aap~382, 152.

\bibitem[Massi et al.(2006)]{Massi2006}
Massi, M.~ et al.~2006, \aap~453, 959.

\bibitem[Massi et al.(2008)]{Massi2008}
Massi, M.~ et al.~2006, \aap~480, 489.


\bibitem[McLean et al.(2000)]{McLean2000}
McLean, I.~et al.~2000, Proc~SPIE~4008, 1048.







\bibitem[O'Neal et al.(1990)]{O'Neal1990}
O'Neal, D.~et al.~1990, \aj~100, 1610.


\bibitem[Onishi et al.(1998)]{Onishi1998}
Onishi, T.~et al.~1998, \apj~502, 296.

\bibitem[Padgett(1996)]{Padgett1996}
Padgett, D.~1996, \apj~471, 847.

\bibitem[Palla \& Stahler(1999)]{Palla1999}
 Palla, F.~\& Stahler, S.~1999, \apj~525, 772.

\bibitem[Palla \& Stahler(2001)]{Palla2001}
 Palla, F.~\& Stahler, S.~2001, \apj~553, 299.

\bibitem[Palla \& Stahler(2002)]{Palla2002}
 Palla, F.~\& Stahler, S.~2002, \apj~581, 1194.

\bibitem[Pichardo et al.(2005)]{Pichardo2005}
Pichardo, B.~et al.~2005, \mnras~359, 521.

\bibitem[Pichardo et al.(2008)]{Pichardo2008}
Pichardo, B.~et al.~2008, \mnras~391, 815.

\bibitem[Pickles(1998)]{Pickles1998}
Pickles, A.~1998, \pasp~110, 863.

\bibitem[Phillips et al.(1996)]{Phillips1996}
Phillips, R.~et al.~1996, \aj~111, 918.

\bibitem[Prato et al.(2001)]{Prato2001}
Prato, L.~et al.~2001, \apj~549, 590.

\bibitem[Prusti et al.(1992)]{Prusti1992}
Prusti, T.~et al.~1992, \aap~259, 537.


\bibitem[Raghavan et al.(2010)]{Raghavan2010}
Raghavan, D.~et al.~2010, \apjs~190, 1.

\bibitem[Rayner et al.(2009)]{Rayner2009}
Rayner, J.~Cushing, M., \& Vacca, W.~2009, \apjs~185, 289.

\bibitem[Rousset et al.(2003)]{Rousset2003}
Rousset, G.~et al.~2003, Proc.~SPIE~4839, 140.

\bibitem[Rydgren et al.(1976)]{Rydgren1976}
Rydgren, A.~Strom, S., \& Strom, K.~1976, \apjs~30, 307.

\bibitem[Siess et al.(2000)]{Siess2000}
 Siess L., Dufour E., \& Forestini M. 2000, \aap~358, 593.

\bibitem[Stefanik et al..(1999)]{Stefanik:99}
 Stefanik, R.\ P., Latham, D.\ W., \& Torres, G.~1999, in Precise
Stellar Radial Velocities, IAU Coll.\ 170, ASP Conf.\ Ser., 185, eds.\
J.\ B.\ Hearnshaw \& C.\ D.\ Scarfe (San Francisco: ASP), 354.




\bibitem[Tamazian et al.(2002)]{Tamazian2002}
Tamazian, V.~et al..~2002, \apj~578, 925 (T2002).

\bibitem[Tokovinin(2008)]{Tokovinin2008}
 Tokovinin, A.~2008, \mnras~389, 925.

\bibitem[Torres et al.(1997)]{Torres1997}
 Torres, G.~et al.~1997, \aj, 114, 2764.

\bibitem[Torres et al.(2002)]{Torres2002}
 Torres, G.~et al.~2002, \aj~124, 1716.

\bibitem[Torres et al..(2003)]{Torres2003}
 Torres, G. et al.~2003, \aj~125, 825.

\bibitem[Torres et al..(2011)]{Torres2011}
 Torres, R.~et al.~2011, \apj~submitted (T2011).

\bibitem[Verrier \& Evans(2007)]{Verrier2007}
Verrier, P.~\& Evans, N.~2007, \mnras~390, 1377.

\bibitem[Wallace \& Hinkle(1996)]{Wallace1996}
 Wallace, L.~\& Hinkle, K.~1996, \apjs~107, 312.


\bibitem[Walker \& Walstencroft(1988)]{Walker1988}
 Walker, H.~\& Walstencroft, R.~1988, \pasp~100, 1509.


\bibitem[Webb et al.(1999)]{Webb1999}
 Webb, R.~et al.~1999, \apj~512, L63.


\bibitem[Welty(1995)]{Welty1995}
Welty, A.~1995, \aj~110, 776 (W1995).

\bibitem[White \& Ghez(2001)]{White2001}
White, R.~\& Ghez, A.~2001, \apj~556, 265.

\bibitem[Wizinowich et al.(2000)]{Wizinowich2000}
Wizinowich, P.~et al.~2000, Proc.~SPIE~4007, 2.

\bibitem[Woitas(2003)]{Woitas2003}
Woitas, J.~2003, \aap~406, 685.

\end{thebibliography}
\end{document}